\documentclass[letterpaper]{article}

\usepackage[T1]{fontenc}

\usepackage{geometry}
\usepackage{setspace}

\usepackage{graphicx}
\usepackage{float}

\usepackage{hyperref}
\usepackage{xcolor}
\usepackage{booktabs}
\usepackage{siunitx}

\newcommand{\Kd}{K\textsubscript{d}}
\newcommand{\Ki}{K\textsubscript{i}}
\newcommand{\IC}[1]{IC\textsubscript{#1}}
\newcommand{\EC}[1]{EC\textsubscript{#1}}

\usepackage{authblk}
\author[1,2]{Geoffrey J.\ Taghon*}
\affil[1]{Johns Hopkins University Whiting School of Engineering, Baltimore, MD, USA}
\affil[2]{National Institute of Standards and Technology, Gaithersburg, MD, USA}

\title{AptaFind: A Lightweight Local Interface for Automated Aptamer Curation from Scientific Literature}
\date{*Email: geoffrey.taghon@nist.gov}

\begin{document}

\maketitle

\begin{abstract}
Aptamer researchers face a literature landscape scattered across publications, supplements, and databases, with each search consuming hours that could be spent at the bench. AptaFind transforms this navigation problem through a three-tier intelligence architecture that recognizes research mining is a spectrum, not a binary success or failure. The system delivers direct sequence extraction when possible, curated research leads when extraction fails, and exhaustive literature discovery for additional confidence. By combining local language models for semantic understanding with deterministic algorithms for reliability, AptaFind operates without cloud dependencies or subscription barriers. Validation across 300 University of Texas Aptamer Database targets demonstrates 84~\% with some literature found, 84~\% with curated research leads, and 79~\% with a direct sequence extraction, at a laptop-compute rate of over 900 targets an hour. The platform proves that even when direct sequence extraction fails, automation can still deliver the actionable intelligence researchers need by rapidly narrowing the search to high quality references.
\end{abstract}

\section*{Keywords}
aptamer, literature mining, artificial intelligence, language model, biocuration, automation


\section{Introduction}

Aptamers are nucleic acid sequences that fold into precise conformations to bind target ligands. These are used for therapeutic and biosensor applications because they offer distinct advantages over proteins: smaller size, higher thermostability, and rapid chemical synthesis.\cite{Cheng2025,Dunn2017,Keefe2010} Yet, discovering relevant aptamer literature resembles navigation with scattered charts rather than consulting a unified map. Unlike protein researchers who turn to UniProt, aptamer investigators must manually traverse publications, supplements, and specialized databases, each search consuming hours depending on database coverage and access restrictions.

The challenge compounds for systematic investigations. Manual databases like the University of Texas Aptamer Database (UTdb) provide high-quality curated data but cannot keep pace with rapidly growing literature.\cite{Askari2024} Commercial literature mining tools (i.e.\ Perplexity) exist but impose cloud dependencies, subscription costs, or proprietary constraints. Simple automated keyword searches yield high false positive rates, missing the contextual information necessary to distinguish aptamers from primers, PCR products, or experimental artifacts. For researchers investigating multiple targets or seeking comprehensive coverage, manual approaches become untenable.

To confront these obstacles, I apply modern automation techniques, with a combination of a language model (LM) and algorithmic processing. LMs excel at contextual extraction from plaintext and restructuring data, precisely the capabilities needed here.\cite{DSouza2025} Sequences appear in multiple nonharmonized places, including main text, tables, and supplementary PDFs. Distinguishing signal from noise requires domain understanding; generic NLP tools that are designed to parse syntactic structures fail on aptamer-specific nomenclature and contexts. Experimental conditions are sometimes reported as prose rather than structured metadata fields. Happily, LMs can now run comfortably on modest laptops and desktops, which enables a new paradigm of on-device, no cost, private processing. However, LMs are not a ``magic bullet'' for sequence extraction. They carry the persistent risk of silent hallucination, which demands monitoring and careful engineering to minimize.\cite{Xu2024} Our solution lies in leveraging the complementary strengths of LMs and hardcoded parsers by introducing the ``Minimum Agentic Flow'' (MAF) principle: contextually-aware LMs for semantic understanding, traditional regular expression (regex) algorithms for data manipulation and validation.

To return the most actionable intelligence per search query, AptaFind implements three-tier research intelligence. Tier~1 results are sequences extracted directly into a local SQL database when possible, the highest value outcome. Tier~2 results consist of curated research leads with complete metadata when extraction fails. This allows users to follow up manually and find the sequences and/or add the downloaded publication into the local AptaFind PDF folder. Tier~3 are the broadest, most general hits: the search was exhaustive for open access documents, but no aptamer-related sequences or literature were found.

This tiered approach recognizes that research value exists along a spectrum. Combined with the MAF principle, AptaFind delivers sensitivity and reliability without cloud costs or external dependencies beyond an internet connection. The platform operates locally using a one billion parameter (Llama3.2 default, $\sim$5\,GB) model, ensuring data privacy and independence from service availability. Validation across three sets of 100 random UT Database targets demonstrates 79~\%--84~\% coverage across all three result tiers, in about 16 min. Our software is presented open source and freely available at \url{https://github.com/usnistgov/aptafind}, to help researchers convert scattered literature into structured knowledge for aptamer discovery.

\section{System Design}

AptaFind's architecture reflects a fundamental insight: research intelligence value exists along a spectrum, not as a binary outcome (Figure~\ref{fig:architecture}). The system orchestrates multi-source literature discovery, enhanced XML processing, and three-tier output generation to ensure every search delivers actionable value.

\subsection{Multi-Source Literature Discovery}

The foundation begins with comprehensive searching. First, any local PDFs are converted to a structured SQL database containing bibliographic and sequence data, if extracted. Found sequences are cross-checked for the correct target assignments with the LM, using a specialized prompt plus the nearby publication context and the paper's abstract. These Tier~1 results form the core of the user's home database, which is the first priority for searches. Next, AptaFind proceeds online to find more information, if desired. First, PubMed queries are created: sequence-specific terms (``5'-'', ``oligonucleotide[Title/Abstract]'') for papers likely containing extractable sequences, and broader fallback queries for exhaustive Tier~3 results. Then, PMC full-text searches attempt to acquire any available full text and supplement documents. Supplement harvesting employs HTML pattern matching to identify links for PDFs, supplemental materials, and supporting information scattered across publisher sites. Because some supplemental files are protected by anti-automation splash pages, Playwright browser automation provides silent file retrieval by simulating a user, with graceful degradation when unavailable. To respect the source server bandwidth, all searches are rate limited (i.e., 3 requests/second for NCBI) and implement exponential backoff for error handling.

\subsection{Enhanced XML Processing and Sequence Extraction}

Once online data is found, it must be structured for further processing. The retrieved content flows through deterministic processing pipelines. Structured metadata extraction uses Python parsing of PubMed XML responses, preserving complete provenance: PMID, DOI, title, authors, journal information. Sequence detection employs pattern matching for 20 to 100 character nucleotide sequences (user-adjustable), containing only valid characters (A/C/G/T/U). I find that context matters profoundly at this step, and this observation was the driver for creating the MAF. When using these nucleotide sequence regex patterns, AptaFind captures primers and PCR products alongside true aptamers. The local language model analyzes the paragraphs surrounding the found sequence with semantic understanding, to distinguish unrelated nucleotide reagents from functional binding sequences. Binding affinity extraction targets \Kd{} and \Ki{} values with unit preservation, again leveraging language models only for locating relevant values while traditional regex handles format harmonization.

\subsection{Deterministic Validation and Deduplication}

Extracted sequences undergo rigorous validation. Nucleotide composition must be valid, length must fall within 20--100 nucleotides, GC content within 20~\%--80~\%. Sequences outside these ranges are flagged for manual review rather than silently discarded, so the system provides intelligence, not simply filtering. Deduplication employs 100~\% identity thresholds across sources, so true single-base sequence variants are preserved. Standardization enforces 5$'$ to 3$'$ orientation, uppercase notation, and parsed modification syntax, if detected (e.g., chemical modifications *dT, \^{}FAM). Through MAF, AptaFind increases reliability, as there are fewer opportunities for the LM tokenizer or output to mutate a sequence. Then, AptaFind proceeds to regex operations, which are fully predictable when applied to harmonized data.

\subsection{Three-Tier Output Generation}

\begin{figure}[htbp]
  \centering
  \includegraphics[width=\textwidth]{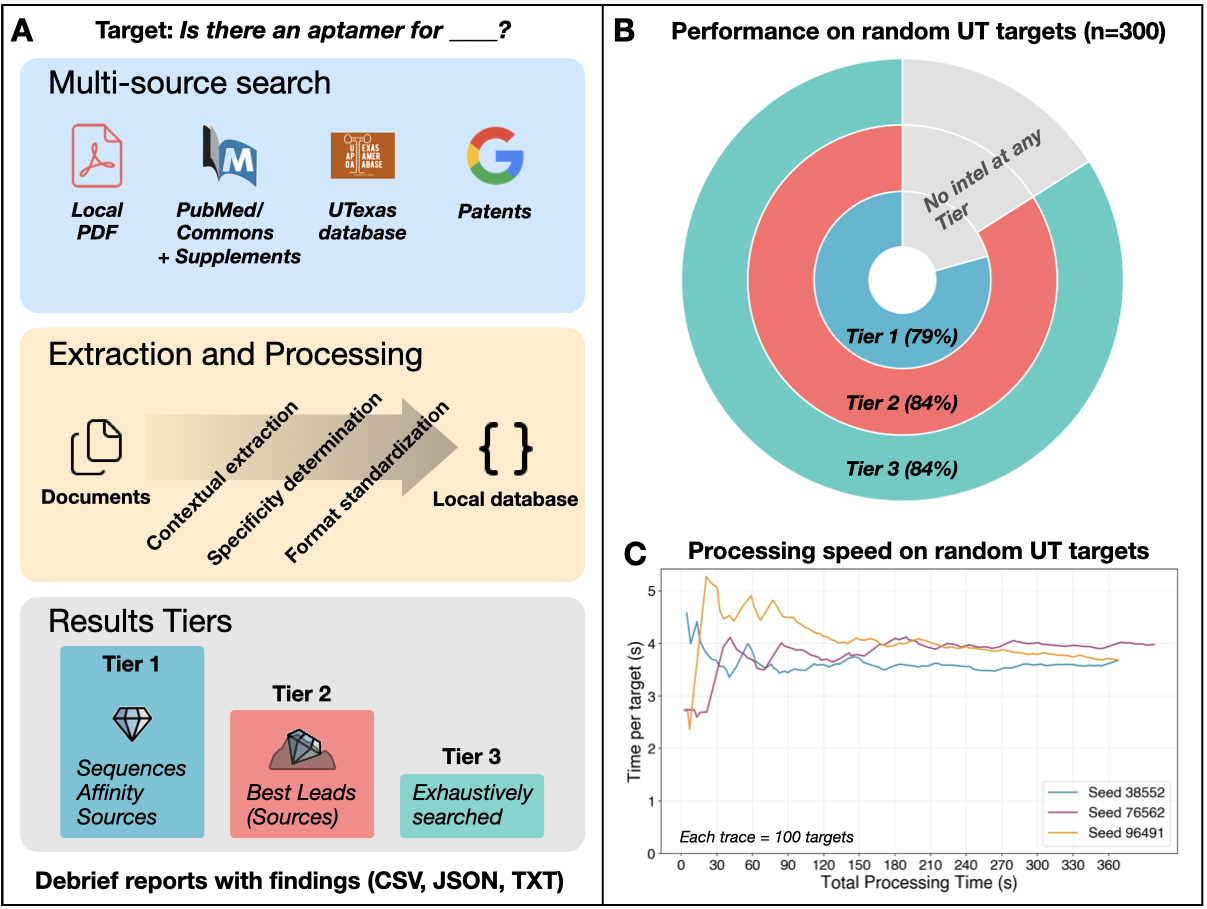}
  \caption{\textbf{AptaFind implements a three-tier research intelligence approach ensuring every search delivers actionable value.} \textbf{(A)} System Architecture: Multi-source literature discovery combines PubMed, PMC, and bioRxiv searches with supplement harvesting and browser automation for comprehensive coverage. Enhanced XML processing extracts structured metadata while deterministic sequence validation ensures reliability. The three tiers deliver calibrated intelligence: Tier~1 (direct sequence extraction, ``diamond quality''), Tier~2 (curated research leads with complete metadata, ``diamonds in the rough''), and Tier~3 (searched exhaustively but found no leads with metadata). Testing three random subsets of 100 targets from the UT Database on a consumer desktop demonstrates \textbf{(B)} processing speed of $\sim$953 targets per hour ($\sim$4 seconds/target) and \textbf{(C)} hit rates across all tiers: $\sim$84~\% literature discovery, $\sim$84~\% research lead generation, 79~\% direct sequences returned.}
  \label{fig:architecture}
\end{figure}

The system delivers three distinct value tiers, from best to least data quality (Figure \ref{fig:architecture}A):

\textbf{Tier 1 (Direct Extraction)} means actual nucleotide sequences (probable aptamers) were found, and updates a local database with sequences, binding affinities, experimental conditions, and complete source attribution. These entries provide immediate utility for researchers seeking known binders.

\textbf{Tier 2 (Research Leads)} means that relevant references with rich metadata were found, but no sequences could be extracted from the text. These are reported with their metadata including direct hyperlinks to their full texts. Thus, if AptaFind can't locate or extract the specific nucleotide sequences, users can manually follow-up with a single click for most queries.

\textbf{Tier 3 (Literature Discovery)} is the lowest tier and means that an exhaustive search identified no aptamer-related literature for the target. The results provided are every unique literature source found for the target. Once Tier~3 has been completed for a target, users should have total, current literature awareness. This is ideal for systematic reviews.

For ease of use, AptaFind offers an intuitive Python GUI application, as well as a CLI for scripting. Processing completes locally without external dependencies or cloud compute costs, typically handling $\sim$1000 target queries per hour on a Mac Studio (Figure \ref{fig:architecture}C). This represents an enormous time savings versus manual literature discovery; assembling the full text set of UTdb references and supplemental documents needed for validation was a 48 hour process for two researchers. Notably, a small subset of the UT-listed DOIs were still inaccessible for NIST employees/academics due to paywalls.

\section{Results and Discussion}

\subsection{Benchmarking}

Benchmarking across three random 100 target sets of the UTdb reveals how the three-tier approach delivers value across the research spectrum (Figure~\ref{fig:architecture}B). Literature discovery (Tier~3) achieved $84.0~\% \pm 3.5~\%$ (SD) coverage, confirming AptaFind's automated search power is comparable to manually curated databases. Research lead generation (Tier~2) likewise reached $84.0~\% \pm 3.5~\%$ (SD), demonstrating the system's ability to transform inaccessible content into actionable intelligence. Direct sequence extraction (Tier~1) succeeded for $79.3~\% \pm 0.6~\%$ (SD) of targets; notable given the diverse access restrictions, format challenges, and paywalls encountered across the corpus. Fast processing speed demonstrates practical utility for systematic screening. The test system (Apple Mac Studio, M2 Max) completes searches at $954 \pm 43$ (SD) targets per hour, so a complete search campaign could be run on large query sets overnight. This throughput enables comprehensive aptamer landscape analysis previously impractical for individual researchers or small laboratories.

Further, I found that the Minimum Agentic Flow principle proves highly useful for output reliability, compared to earlier LM-centric or regex-centrioc attempts. Language models handle semantic understanding (context-aware sequence identification, binding affinity, target name harmonization) and context that regex cannot detect, while deterministic algorithms ensure validation, deduplication, and formatting. This division of labor leverages model strengths while mitigating hallucination risk through traditional algorithms for all data transformation \cite{Reche2025,Alecci2025}.

\subsection{Limitations and Future Directions}

Paywall-protected content remains inaccessible to the code, though the research lead generation (Tier~2) partially mitigates this by providing complete reference metadata. Users can also manually download any desired publications and add them to the SQL database, circumventing the problem. Reference matching between main text and supplemental files is dependent on shared context or keywords between the documents; currently, the pipeline treats each PDF file as a separate searchable entity. This can be mitigated by manually concatenating main and supplemental texts into a single document. Sequence extraction from images is not yet implemented, and complex tables require detection enhancement, a tractable problem as multimodal/vision language models mature. Binding affinity extraction succeeds when values appear as harmonized \Kd{}/\Ki{} measurements but struggles with qualitative descriptions (``less than'', ``100x reduced'') or unconventional units like \EC{50}/\IC{50}.

\subsection{Broader Impact}

AptaFind's three-tier intelligence approach extends beyond aptamers. Any scientific research domain facing scattered literature, diverse formats, and access restrictions could benefit from this framework: synthetic biology parts, small molecule binders, engineered proteins. The MAF principle (language models for semantic understanding, algorithms for reliability) provides a template for reliable agentic systems in data-intensive research. As genome-scale biomolecular engineering accelerates, the ability to rapidly survey existing literature becomes increasingly critical. AptaFind demonstrates that local, lightweight tools can deliver research intelligence calibrated to actual accessibility constraints rather than treating literature mining as binary success. We must embrace the messy reality of thousands of different reporting formats, because the goal is not perfect extraction but actionable intelligence for the journey ahead.

\section{Methods}

For the most detailed description of the process, see the PSEUDOCODE file in the GitHub repository.

\subsection{Database Construction from PDFs}

In brief: First, PDFs in the user-specified directory are converted to markdown. Next, semantic chunks of 1000 tokens are segmented, to preserve context. Section metadata is labeled, such as abstract, introduction, etc. After conversion, a ColBERT search index is built using RAGatouille, an open source Retrieval Augmented Generation (RAG) pipeline. Then, nucleotide sequences and experimental data are extracted using regex patterns. AptaFind extracts DNA/RNA sequences (20--100 nucleotides), binding affinities (\Kd{}, \IC{50}, \EC{50}) with unit conversion, and experimental metadata (pH, temperature, buffer). The next phase is LM-guided curation. With a series of prompts, the LM attempts target verification of extracted sequences and confirmation that the metadata makes sense. The LM outputs a result confidence score, based on its ``own judgement'' of the sequence and context. The result is a searchable sequence database with source attribution to the original PDF.

\subsection{Online Literature Search Implementation}

In brief: First, PubMed queries use the Entrez E-utilities API with multi-tier search strategies. Sequence-specific queries prioritize papers likely to contain extractable sequences using terms like ``sequence[Title/Abstract]'', ``5'-[Title/Abstract]'', ``oligonucleotide[Title/Abstract]''. Broader fallback queries ensure comprehensive literature coverage when sequence-specific searches yield no results. PMC full-text queries leverage open access availability and supplement harvesting capabilities. bioRxiv searches use web scraping with rate limiting to access preprint literature when API access is limited. All searches respect API rate limits (3 requests/second for NCBI) and implement exponential backoff for error handling. Supplement harvesting employs BeautifulSoup pattern matching to identify PDF links, supplemental materials, and supporting information. Playwright browser automation provides fallback access for anti-bot protected materials with graceful degradation when unavailable.

\subsection{Enhanced XML Processing and Validation}

Article metadata extraction uses xmltodict for structured parsing of PubMed XML responses. Sequence detection employs deterministic pattern matching for nucleotide sequences (20--100 characters, valid nucleotides A/C/G/T/U only). Contextual filtering distinguishes aptamers from primers using surrounding text analysis and experimental section identification. Binding affinity extraction targets \Kd{} and \Ki{} values with unit preservation using regex pattern matching. Source attribution maintains complete provenance with PMID, DOI, title, authors, and journal metadata for traceability.

\subsection{Sequence Processing Algorithms}

I set a 100~\% identity threshold for sequence deduplication across sources. Sequence harmonization rules are as follows: Enforce 5$'$$\rightarrow$3$'$ orientation, convert to uppercase, parse modification notation (e.g., *dT, \^{}FAM). When validation checks occur, nucleotides must be in \{A, C, G, T, U\}, have a length 20--100 characters (bases), and a GC content of 20--80~\%. Sequences outside validation ranges are flagged for manual review as outliers, but not automatically discarded.

\subsection{UT Database Validation Protocol}

I used the UT Aptamer Database snapshot dated September 2023, containing 555 total ligands. For each ligand, I ran AptaFind with default parameters and compared outputs to UTdb ground truth sequences. Per-tier recovery percentage is defined as (sequences with a Tier~$n$ hit) / (100 total queries). Time measurement is reported as the wall-clock time in seconds from query initiation to final report generation.

\subsection{Statistical Analysis}

Results for each metric are reported as the mean success percentage $\pm$ one standard deviation (SD) across three tests (3 $\times$ (n=100)).

\subsection{Software Availability}

Source code: \url{https://github.com/usnistgov/AptaFind}

License: NIST (open source)

\subsection{Hardware Specifications}

Development/validation server: Mac Studio 2023 (Apple M2 Max CPU, 64\,GB RAM, MacOS 15.6.1).


\section*{Acknowledgements}

G.J.T.\ was supported in part by an appointment to a Research Associateship Program at the National Institute of Standards and Technology, administered by the Johns Hopkins University Whiting School of Engineering. Certain tools and software are identified in this paper to foster understanding. Such identification does not imply recommendation or endorsement by the National Institute of Standards and Technology, nor does it imply that the tools and software identified are necessarily the best available for the purpose. The views expressed in this publication are those of the author and do not necessarily represent the views of the U.S.\ Department of Commerce or the National Institute of Standards and Technology. G.J.T. gratefully acknowledges Samuel Schaffter (National Institute of Standards and Technology) for expert consultation about aptamers; the University of Texas at Austin Aptamer Database curators for maintaining high-quality ground truth data; and open access publishers for enabling literature search automation.

\section*{Supporting Information}

The following files are available:
\begin{itemize}
  \item File S1: Complete prompt templates for literature search and sequence extraction
  \item File S2: Example output file set for a ligand (CSV, JSON, TXT formats)
  \item File S3: Pseudocode of the MAF processing workflow for aptamer data: \url{https://github.com/usnistgov/aptafind/blob/main/PSEUDOCODE.md}
\end{itemize}

\bibliographystyle{ieeetr}
\bibliography{aptafind-refs}

\end{document}